\documentclass[prd,draft,amsfonts,amssymb,preprintnumbers,showpacs]{revtex4}
\usepackage{graphicx,graphics,epsf,amsfonts,amssymb,amsbsy}

\newcommand{\tr}{\mathop{\rm tr}\nolimits}

\newcommand{\Tr}{\mathop{\rm Tr}\nolimits}

\newcommand{\e}{\mathop{\rm e}\nolimits}

\newcommand{\ds}{\displaystyle}

\newcommand{\ii}{\rm i}
\newcommand{\Det}{\mathop{\rm Det}\nolimits}

\begin{document}
\preprint{Preprint HU-EP-06/45}
\title{Restoration of Dynamically Broken Chiral and Color Symmetries
  for an Accelerated Observer} 
\author{D.~Ebert$^{1}$ and  V.~Ch.~Zhukovsky$^{2}$}
\affiliation{$^{1}$ Institut f\"ur Physik,
Humboldt-Universit\"at zu Berlin, 12489 Berlin, Germany}
\affiliation{$^{2}$ Faculty of
Physics, Department of Theoretical Physics, Moscow State University,
119899, Moscow, Russia
}

\date{\today}
\newcommand{\bmx}{\mbox{\boldmath $x$}}
\newcommand{\bmy}{\mbox{\boldmath $y$}}
\newcommand{\bmk}{\mbox{\boldmath $k$}}
\newcommand{\bmp}{\mbox{\boldmath $p$}}
\newcommand{\bmq}{\mbox{\boldmath $q$}}
\newcommand{\bmP}{\mbox{\boldmath $P$}}  
\newcommand{\kfey}{\ooalign{\hfil/\hfil\crcr$k$}}
\newcommand{\pfey}{\ooalign{\hfil/\hfil\crcr$p$}}
\newcommand{\qfey}{\ooalign{\hfil/\hfil\crcr$q$}}
\newcommand{\Deltafey}{\ooalign{\hfil/\hfil\crcr$\Delta$}}
\newcommand{\nablafey}{\ooalign{\hfil/\hfil\crcr$\nabla$}}
\newcommand{\Dfey}{\ooalign{\hfil/\hfil\crcr$D$}}
\newcommand{\partfey}{\ooalign{\hfil/\hfil\crcr$\partial$}}
\def\sech{\mathop{\rm sech}\nolimits}

\begin{abstract}
We study the behavior of quark and diquark condensates  at finite Unruh
temperature as seen by an
accelerated observer. The gap equations for these condensates have been obtained
with consideration of a finite chemical potential. Critical values
of the acceleration for the restoration of chiral and color symmetries have been
estimated. 
\end{abstract}


\maketitle

\section{Introduction}

It is well known that, according to the Hawking--Unruh
effect~\cite{hawking,8a}, an observer accelerated uniformly in the QCD 
vacuum behaves as if he were in a thermal bath at the Unruh
temperature $T_{U}=a/2\pi$ ($a$ is an acceleration constant). For
interacting field theories this is demonstrated by the fact that
Euclidean Green's functions 
written in terms of the Rindler coordinates 
are periodic in time, 
and therefore they may be interpreted as thermal. 
Several problems have been studied in relation to the Hawking-Unruh
effect: see, for instance, \cite{lau} concerning the discussion of a uniformly
accelerated oscillator, or 
\cite{fedotov} with the study of pair creation by a 
homogeneous electric field from the point of view of an accelerated
observer.
At the same time, in a recent paper  
\cite{glass}, it was argued that the phase transition from a color glass
condensate to a quark gluon plasma through the mechanism of the Hawking-Unruh thermalization
can become experimentally observable in relativistic heavy ion collisions. 

It was proposed more than twenty years 
ago \cite{Ba,Frau,bl} that at high baryon densities a colored
diquark condensate $<qq>$ might appear.
In analogy with the ordinary
superconductivity, this effect was called color superconductivity
(CSC). 
In particular, the CSC phenomenon was investigated in the framework of the one-gluon
exchange approximation in QCD \cite{son}, where
the colored Cooper pair formation is predicted selfconsistently
at extremely high values of the chemical potential $\mu\gtrsim 10^8$
MeV \cite{raj}. Unfortunately, such baryon densities are not
observable in nature and not accessible in experiments (the typical
densities inside the neutron stars or in the future heavy ion
experiments correspond to  $\mu\sim 500$ MeV). In order to study the
problem at lower values of $\mu$, various effective 
theories for low energy QCD, such as the instanton model \cite{rapp}
and the Nambu$-$Jona-Lasinio (NJL) model\cite{nambu} can be employed. 

It is well known that effective field theories with four-fermion
interaction (the so-called Nambu -- Jona-Lasinio (NJL) models), which
incorporate the phenomenon of dynamical chiral symmetry breaking, are
quite useful in describing  low-energy hadronic processes (see
e.g. \cite{7,kunihiro} and references therein).
Since the NJL model displays the same symmetries as QCD,
it can be successfully used for simulating some of the QCD ground
state properties under the influence of external conditions such as
temperature, baryonic chemical potential, or even curved
spacetime \cite{kunihiro,buballa,odin,klev,klim}. In
particular, the role of the NJL approach increases,
when detailed numerical lattice calculations are not yet admissible in
QCD with nonzero baryon density and in the
presence of external gauge fields \cite{kl,ekvv,khud}.

The
possibility for the existency of the CSC phase in the region
of moderate densities was recently proved 
(see, e.g., the papers \cite{rapp,klev,alf,berg}
as well as the review articles \cite{alf2} and references therein). 
In these papers it was shown 
that the diquark condensate $<qq>$ can appear  already at a rather moderate
baryon density ($\mu\sim 400$ MeV). The conditions favorable for this
condensate to be formed can possibly  exist in the cores of cold neutron stars. 
Since quark Cooper pairing occurs in the color anti-triplet
channel, a nonzero value of $<qq>$ means that, apart from
the electromagnetic $U(1)$ symmetry,  the color
$SU_c(3)$ symmetry should be spontaneously broken  inside
the CSC phase as well. 
In the framework of NJL models the CSC phase formation has generally 
been considered
as a dynamical competition between diquark  $<qq>$ and
usual quark-antiquark condensation $<\bar qq>$.

 Recently,  the
dynamical chiral symmetry breaking and its restoration 
for a uniformly accelerated observer  due to the thermalization effect  
of acceleration was studied in~\cite{ohsaku2} at zero chemical
potential.  Further investigations of 
the possible influence 
of the Unruh temperature on the phase transitions in dense quark
matter with a finite chemical potential, and especially on the restoration
of the broken color symmetry in CSC is thus especially 
interesting.
Related problems
have also been studied for chiral symmetry breaking in curved spacetime
\cite{odin,huang}, which may be useful 
for the investigation of compact stars, where the gravitational field
is strong and its effect cannot be neglected. 
Obviously, the  results of these studies might have some relevance to
the physics of black holes, whose surface gravity causes
the finite temperature effects (the Hawking effect). The Rindler
metric may be regarded here as an approximation of the situation near
the event horizon of a black hole. 
Thus, the study of phase transitions in quark  
matter with quark and diquark condensates under the influence of
acceleration or, equivalently, 
strong gravitational fields 
is of great interest. 

In this paper, we study  quark and diquark condensates as functions
of the Unruh temperature and finite chemical potential by using a
NJL-type model formulated in Rindler coordinates.
As our main result, from the gap equations for these condensates written in
Rindler coordinates, the critical values of acceleration (the critical Hawking-Unruh
temperatures) for the restoration of the broken chiral and color symmetries were 
obtained. The results exactly coincide with the usual temperature
restoration of these symmetries, when the corresponding relation
between the acceleration and the Unruh temperature is taken into account. 

\section{Nambu--Jona-Lasinio model in Rindler coordinates}

The required effective NJL field theory in Rindler coordinates will be
obtained by a suitable transformation of a NJL-type model in flat
spacetime with Minkowski coordinates
$(x^{0},x^{1},\vec x_{\perp})$. For this aim, let us first quote some 
useful definitions and formulas.
The physics for an
accelerated observer can be described by transforming to the Rindler coordinates
$(\eta,\rho,\vec x_{\perp})$ by means of the following coordinate
transformation:
\[
x^{0}=\rho\sinh a\eta, \quad x^{1}=\rho\cosh a\eta, \quad x^i=x^i \quad (i=2,3),
\] 
defined on the right Rindler wedge,
\[ 
0<\rho<+\infty, \quad -\infty<\eta<+\infty,
\]
and on the left Rindler wedge,
\[
-\infty<\rho<0, \quad -\infty<\eta<+\infty,
\]
where $\eta$ is the time variable in Rindler coordinates.

The gamma-matrices $\gamma_{\mu}$, the metric $g_{\mu\nu}$ and the
vierbein $e^{\mu}_{\hat{a}}$, as well as the definitions of the covariant
derivative $\nabla_{\nu}$ and  
spin connection $\omega^{\hat{a}\hat{b}}_{\nu}$ are given by the following
relations~\cite{brill}: 
\begin{eqnarray}
& & \{\gamma_{\mu}(x),\gamma_{\nu}(x)\}=2g_{\mu\nu}(x), \quad
  \{\gamma_{\hat{a}},\gamma_{\hat{b}}\}=2\eta_{\hat{a}\hat{b}}, \quad
  \eta_{\hat{a}\hat{b}}={\rm diag}(1,-1,-1,-1), \nonumber \\ 
& & g_{\mu\nu}g^{\nu\rho}=\delta^{\rho}_{\mu}, \quad
  g^{\mu\nu}(x)=e^{\mu}_{\hat{a}}(x)e^{\nu \hat{a}}(x), \quad
  \gamma_{\mu}(x)=e^{\hat{a}}_{\mu}(x)\gamma_{\hat{a}}. \label{2}\\
 & &
  \nabla_{\nu}\equiv\partial_{\nu}+\frac{{1}}{2}\sigma_{\hat{a}\hat{b}}
\omega^{\hat{a}\hat{b}}_{\nu}, 
  \quad
  \sigma_{\hat{a}\hat{b}}\equiv\frac{{1}}{4}[\gamma_{\hat{a}},\gamma_{\hat{b}}],
  \nonumber \\  
& &
  \omega^{\hat{a}\hat{b}}_{\mu}\equiv\frac{1}{2}e^{\hat{a}\lambda}
e^{\hat{b}\rho}[C_{\lambda\rho\mu}-C_{\rho\lambda\mu}-C_{\mu\lambda\rho}],
  \quad C_{\lambda\rho\mu}\equiv
e^{\hat{a}}_{\lambda}\partial_{[\rho}e_{\mu]\hat{a}}
\label{3}
\end{eqnarray}
Here, the index $\hat{a}$ refers to the flat tangent space defined by the
vierbein at spacetime point $x$, and the $\gamma^{\hat{a}}
(a=0,1,2,3)$ are the usual Dirac gamma-matrices  of Minkowski spacetime. 

The line element 
\[
ds^2=\eta_{\hat a\hat b}e_\mu^{\hat a}e_\nu^{\hat b}dx^\mu dx^\nu
\]
in these coordinates with the vierbeins
\[
\e_0^{\hat0}=a\rho,\quad e_1^{\hat 1}=\dots=1
\]
is given by the relation
\[
ds^2=a^2\rho^2d\eta^2-d\rho^2-d\vec x_{\perp}^2
\]
with the metric tensor
\begin{eqnarray}
g_{\mu\nu}=(a^2\rho^2,-1,-1,-1).
\label{position1}
\end{eqnarray}
In what follows, we shall limit our consideration to the right Rindler
wedge. An observer at fixed $\rho,\vec x_{\perp}$ measures a proper time
$d\tau=a\rho d\eta$ and has a proper acceleration $1/\rho$. The observer at
$\rho=1/a$ measures $d\tau=d\eta$ and has a proper acceleration $a$. 
The world line of the observer in Rindler coordinates is thus given as
\begin{eqnarray}
\eta(\tau)=\tau, \quad \rho(\tau)=1/ a, \quad \vec x_{\perp}(\tau)={\rm
  const.}
\label{position}
\end{eqnarray}

For an accelerated observer moving with constant acceleration
according to (\ref{position}), the
gamma matrices in Rindler coordinates are obtained by the definition
given in (\ref{2})
\begin{eqnarray}
\gamma_0=a^2\rho^2\gamma^0=a\rho\gamma_{\hat 0},\quad
\gamma_\mu=g_{\mu\nu}\gamma^\nu, 
\label{position2}
\end{eqnarray}
 and hence
\begin{eqnarray}
\gamma^{0}(x)=\frac{1}{a\rho}\gamma^{\hat{0}}, \quad
\gamma^{1}(x)=\gamma^{\hat{1}}, \quad \gamma^{2}(x)=\gamma^{\hat{2}},
\quad \gamma^{3}(x)=\gamma^{\hat{3}}.  
\end{eqnarray}

Using the definition given in (\ref{3}) for computing the spin connection, 
one finds the components of the covariant derivatives in Rindler coordinates: 
\begin{eqnarray}
\nabla_{0}=\partial_{\eta}+\frac{a}{2}\gamma_{\hat{0}}\gamma_{\hat{1}},
\quad \nabla_{1}=\partial_{\rho}, \quad \nabla_{2}=\partial_{2},
\quad \nabla_{3}=\partial_{3}.   
\end{eqnarray}

After these intoductions, let us consider the following four-quark
model of up- and down-quarks with $(\bar q q)$ and $(q q)$
interactions in the color group $SU_c(N_c)$ given by the
generalization of the corresponding  
Lagrangian for the Nambu--Jona-Lasinio model  in flat
space \cite{berg,khud}, \footnote{The most general four-fermion interaction would 
include additional vector and axial-vector $(\bar q q)$ as
well as pseudo-scalar, vector and axial-vector-like $(q q)$
-interactions. 
For our goal of studying the effect of acceleration 
on the competition of quark and diquark condensates, the
interaction structure of (\ref{x1}) is, however, sufficiently general.}
\begin{eqnarray}
 {\mathcal L}&=&\bar q(x)[{\rm i}\gamma^\nu(x)\nabla_\nu +
 \mu\gamma^0(x)]q+\frac{G_1}{2N_c}[(\bar q(x)q(x))^2+(\bar q(x){\rm i}\gamma^5\vec
  \tau q(x))^2]+\nonumber\\
  &+&\frac{G_2}{N_c}\sum_b[{\rm i}\bar q(x)_c\varepsilon
({\rm i}\lambda_{as}^b)       \gamma^5 q(x)]
  [{\rm i}\bar q(x)\varepsilon
({\rm i}\lambda_{as}^b)\gamma^5 q_c(x)].
\label{x1}
\end{eqnarray}
In (\ref{x1})  $\mu$ is the quark chemical potential, $q_c=C\bar
q^t$, $\bar q_c=q^t C$ are charge-conjugated spinors,
and $C={\rm i}\gamma^2\gamma^0$ 
is the charge conjugation matrix ($t$ denotes
the transposition operation). It is necessary to note that in order to obtain 
realistic estimates for masses of 
vector/axial-vector mesons and diquarks 
in extended NJL--type of models \cite{7,ebertt}, we have to allow for
independent coupling constants 
$G_1, G_2$, rather than to consider them related by a Fierz
transformation
of a current-current interaction via gluon exchange.
In what follows we assume $N_c=3$ 
and replace the antisymmetric color matrices
$\lambda_{as}^b$ (with a factor ${\rm i}$) by the antisymmetric
$\epsilon^b$
operator. The quark field $q\equiv q_{i\alpha}$ is
a flavor doublet and color triplet as well as a four-component Dirac
spinor, where $i=1,2$; $\alpha = 1,2,3$. (Latin and Greek indices
refer to flavor and color 
indices, respectively; spinor indices are
omitted.) Furthermore,  $\vec \tau\equiv (\tau^{1},
\tau^{2},\tau^3)$ are Pauli
matrices in the flavor space; $(\varepsilon)^{ik}\equiv\varepsilon^{ik}$,
$(\epsilon^b)^{\alpha\beta}\equiv\epsilon^{\alpha\beta b}$
are totally antisymmetric tensors in the flavor and color spaces,
respectively. Clearly, the Lagrangian (\ref{x1}) is invariant under
the chiral $SU(2)_L\times SU(2)_R$ and color $SU_c(3)$ groups.

Next, by applying the usual bosonization procedure, we obtain the
linearized version of the model (\ref{x1}) with collective bosonic 
fields 
\begin{eqnarray}
\tilde {\cal L}\ds &=&\bar q[{\rm i}\gamma^\nu\nabla_\nu+\mu\gamma^0]q
 -\bar q(\sigma+{\rm i}\gamma^5\vec
 \tau\vec\pi)q-\frac{3}{2G_1}(\sigma^2+\vec \pi^2)-
 \nonumber\\
&-&\frac3{G_2}\Delta^{*b}\Delta^b-
\Delta^{*b}[{\rm i}q^tC\varepsilon\epsilon^b\gamma^5 q]
 -\Delta^b[{\rm i}\bar q \varepsilon\epsilon^b\gamma^5 C\bar q^t].
 \label{x2}
\end{eqnarray}
The Lagrangians (\ref{x1}) and (\ref{x2}) are equivalent, as can be
seen by using the
equations of motion for bosonic fields, from which it follows that
\begin{equation}
  \Delta^b\sim {\rm i}q^t C\varepsilon \epsilon^b\gamma^5 q,\quad
  \sigma\sim\bar qq,\quad
  \vec \pi\sim {\rm i}\bar q\gamma^5\vec\tau q.
\end{equation}

Clearly, the $\sigma$ and $\vec\pi$ fields are color singlets.
Besides, the (bosonic)
diquark field $\Delta^{b}$ is a color antitriplet and a (isoscalar)
singlet
under the chiral
$SU(2)_L\times SU(2)_R$ group. Note further that the $\sigma$,
$\Delta^b$, are scalars, but
the $\vec\pi$ are pseudo-scalar fields. Hence, if $\sigma\ne0$, then
chiral
symmetry of the model is spontaneously broken, whereas $\Delta^b\ne0$
indicates the dynamical
breaking of both the color and electromagnetic  symmetries of the
theory. In what follows, for our purpose of investigating the ground
state of the system, we may suppose that all
boson fields  do not depend on space-time.

In the one-loop approximation, the partition function can be written
as follows:
$$
Z\,\,=\,\,  \exp({\rm i} S_{\rm
  eff}(\sigma,\vec\pi,\Delta^b,\Delta^{*b})),
$$
where 
\begin{equation}
  S_{\rm eff}(\sigma,\vec\pi,\Delta^b,\Delta^{*b})=-N_c\int
  d^4x\sqrt{-g}
  \left[\frac{\sigma^2+\vec\pi^2}{2G_1}+\frac{\Delta^b\Delta^{*b}}{G_
  2}\right]+ S_q
  \label{x4}
\end{equation}
is  the effective action for the boson fields. 
The quark contribution to the effective action,  $S_q$, is expressed
through the path integral over quark fields
\begin{equation}
  Z_q=\exp({\rm i} S_q)=N'\int d\bar q dq\exp\Bigl({\rm i}\int d^4
  x\sqrt{-g}\left[ \bar q  D q+
  \bar q{\cal M}\bar q^t+q^t\bar{\cal M}q\right]\Bigr).
\label{5}
\end{equation}
where $N'$ is a normalization constant.

In (\ref{5}) we have used the following notations
\begin{equation}
    D={\rm i}\gamma^\mu\nabla_\mu-\sigma-{\rm i}\gamma^5\vec\pi\vec\tau+\mu\gamma
  ^0,\qquad
  \bar{\cal M}=-{\rm i}\Delta^{*b}C\varepsilon\epsilon^b\gamma^5,\qquad
  {\cal M}=-{\rm i}\Delta^{b}\varepsilon\epsilon^b\gamma^5C,\qquad
\end{equation}
where $D$ is an operator in the coordinate, spinor and flavor
spaces, whereas
 ${\cal M}$ and $\bar{\cal M}$ are operators in the color space as well.
Next,  assume that in the ground state of our model
$\langle\Delta^1\rangle=
\langle\Delta^2\rangle=\langle\vec\pi\rangle=0$ and
$\langle\sigma\rangle$, ${\langle\Delta^3\rangle\ne0}$
\footnote{If
$\langle\vec\pi\rangle\ne0$ then one would have spontaneous breaking
of parity.
For strong interactions parity is, however, a conserved
quantum number, justifying the assumption
$\langle\vec\pi\rangle=0$.}.
Obviously, the residual color symmetry group of such a vacuum
is $SU_c(2)$ whose generators are the first three generators of
the initial $SU_c(3)$.

Due to our assumption on the vacuum structure,
we put $\Delta^{1,2}\to \langle\Delta^{1,2}\rangle\equiv 0$, 
$\vec\pi\to \langle\vec\pi\rangle = 0$, $\langle\Delta^3\rangle\equiv\Delta,\,\,\langle\sigma
\rangle\equiv\sigma$. One can
easily see that the functional integral in (\ref{5})
can be factorized
\begin{eqnarray}
  \label{kx9}
 Z_q=\exp({\rm i} S_q)=N'\int d\bar
  q_3 dq_3\exp\Bigl({\rm i}\int d^4x\sqrt {-g}
  \bar q_3 \tilde D q_3\Bigr)\nonumber\\
  \times \int d  \bar Q dQ\exp\Bigl({\rm i}\int d^4x\sqrt{-g}\left[\bar Q\tilde{ D}Q+\bar
  Q M\bar Q^t+
  Q^t\bar MQ\right]\Bigr),
  \label{x9}
\end{eqnarray}
where  $q_3$ is the quark field of color 3, and 
$Q\equiv(q_1,q_2)^t$ is the doublet, composed from quark fields of
the colors 1,2. Moreover,
\begin{equation}
\tilde D=D|_{\vec\pi=0},\,\,    \bar M=-i\Delta^* C\varepsilon\tilde\epsilon\gamma^5,\quad
  M=-i\Delta\varepsilon\tilde\epsilon\gamma^5 C.
  \label{x10}
\end{equation}
In (\ref{x10}), $\tilde \epsilon$ is the matrix in the two-dimensional
color subspace, corresponding to the $SU_c(2)$ group:
$$
\tilde\epsilon=
\left (\begin{array}{cc}
0 & 1\\
-1 &0
\end{array}\right ).
$$
Clearly, the integration over $q_3$ in (\ref{kx9}) yields
$\Det \tilde D$.

Defining $\Psi^t=(Q^t, \bar{Q})$ and introducing the matrix-valued
operator 
$$
Z=
\left (\begin{array}{cc}
2\bar M~, & -\tilde{ D}^t\\
\tilde{ D}~, &2M
\end{array}\right ),
$$
the Gaussian integral over $\bar Q$ and $Q$ in (\ref{x9}) can be
rewritten in compact matrix notation and be evaluated as 
\begin{equation}
  \int d\Psi  \e^{\ds\frac {\rm i}2\int d^4x\sqrt{-g}\Psi^t Z\Psi}=\sqrt{\Det Z}.
\label{90}
\end{equation}
Then, by using in (\ref{90}) the general formula
$$
\Det\left 
(\begin{array}{cc}
U~, & V\\
\bar V~, & \bar U
\end{array}\right )=\Det \left[-\bar VV+\bar VU\bar V^{-1}\bar U\right]=\Det
\left[\bar UU-\bar UV\bar U^{-1}\bar V\right],
$$
one obtains the result:
\[\exp ({\rm i} S_q) =
N'\Det \tilde D\,\,{\Det}^{1/2}[4M\bar M+ M\tilde D^t
M^{-1}\tilde
D]\]
\begin{equation}
=N'\Det({\rm i}\gamma^\nu\nabla_\nu-\sigma+\mu\gamma^0)
{\Det}^{1/2}\left [4|\Delta|^2
+(-{\rm i}\gamma^\nu\nabla_\nu-\sigma+\mu\gamma^0)({\rm i}\gamma^\mu\nabla_\mu-\sigma+\mu\gamma^0)\right ].
\label{kp3_12}
\end{equation}
Recall that the first $\Det$-operation
in (\ref{kp3_12}) acts only in the flavor,
coordinate and spinor spaces, whereas the second $\Det$-operation acts
in the two-dimensional color subspace, as well. The quark contribution
to the effective action 
\[
S_q=-{\rm i}\Tr \log({\rm
  i}\gamma^\nu\nabla_\nu-\sigma+\mu\gamma^0)-{{\rm i}\over 2}\Tr\log\left [4|\Delta|^2
+(-{\rm i}\gamma^\nu\nabla_\nu-\sigma+\mu\gamma^0)({\rm
  i}\gamma^\mu\nabla_\mu-\sigma+\mu\gamma^0)\right ] 
\]
can be written in the form
\begin{equation}
S_q\,=\,S_{q1}\,+\,S_{q2}=-{{\rm i}\over 2} \left[\tr\log B_1^2\,+\,2\tr\log B_2^2\right ],
\label{Sq}
\end{equation}
where we have summed over colors (leading to the factor 2 in the
second term; the $\tr-$operation does not include color indices any more) and   
\[
B_1^2\,=\,(-{\rm i}\gamma^\nu\nabla_\nu-\sigma-\mu\gamma^0)({\rm
  i}\gamma^\mu\nabla_\mu-\sigma+\mu\gamma^0),\]
\begin{equation}
B_2^2\,=\,4|\Delta|^2+(-{\rm i}\gamma^\nu\nabla_\nu-\sigma+\mu\gamma^0)({\rm
  i}\gamma^\mu\nabla_\mu-\sigma+\mu\gamma^0).
\label{B}
\end{equation}
Here the product of the relevant operators appearing in (\ref{B}) can
be represented in Rindler coordinates as
\begin{equation}
B^2\equiv(-{\rm i}\gamma^\nu\nabla_\nu-\sigma)({\rm
  i}\gamma^\mu\nabla_\mu-\sigma)\,=\,{1\over \rho^2}\left[{1\over
  a}\partial_\eta + {1\over 2}\gamma_{\hat 0}\gamma_{\hat 1}\right
  ]^2-\left [{\partial^2\over \partial \rho^2}+{1\over
  \rho}{\partial\over \partial \rho}-(\vec \gamma \vec
    \nabla)_{\perp}^2-\sigma^2\right ].
\label{sq}
\end{equation}
For the following, it is convenient to represent the operators $B_{1}^2,\,B_{2}^2$ in the basis
of the solutions of the squared Dirac equation 
\begin{equation}
B^2\Psi_{\vec k_{\perp},j}(\eta,\vec x_{\perp},\rho)=0.
\label{dirac}
\end{equation}
The solutions can be sought in the form
\begin{equation}
\Psi_{\vec k_{\perp},j}(\eta,\vec
x_{\perp},\rho)=\e^{-{\ii}aj\eta}\e^{{\ii}\vec k_{\perp}\vec x_{\perp}}
\psi_{j}(\rho),  
\label{solution}
\end{equation}
and hence, with consideration of (\ref{sq}), the function $
\psi_{j}(\rho)$ is the solution of the  second order Bessel 
differential equation forming the basis of the Rindler modes (see,
e.g., \cite{hawking})
\begin{equation}
\left(\rho^2{d^2\over d\rho^2}+\rho{d\over
    d\rho}-m^2\rho^2+E_j^2\right)\psi_{j}(\rho)=0,
\label{bess}
\end{equation}
where for the Rindler modes in the
fermion sector, we have to take 
 \begin{equation}
m^2={\vec k}_{\perp}^2+\sigma^2,\,\, E_j^2=(j\pm{{\rm i}\over 2})^2, \,\,0<j<+\infty.
\label{qnumb}
\end{equation}
Two independent solutions of equations (\ref{dirac}), (\ref{bess}) can be obtained
by using the projection operator $P_{\pm}={1\over 2}(1\pm\gamma_{\hat
    0}\gamma_{\hat 1}):$
\[
\psi^{(+)}_j=P_+\psi_j, \quad \psi^{(-)}_j=P_-\psi_j.
\]   
Then the normalized solutions of (\ref{bess}) look as follows:
\begin{eqnarray}
\psi^{(-)}_{j}(\rho)= \frac{\sqrt{(-2{\rm i}j-1)\cosh\pi
    j}}{\pi}K_{{\rm i}j+\frac{1}{2}}(m\rho), &\quad& \psi^{(+)}_{j}(\rho) =
\frac{\sqrt{(+2{\rm i}j-1)\cosh\pi j}}{\pi} K_{{\rm i}j-\frac{1}{2}}(m \rho), \nonumber \\
& &  
\label{basis}
\end{eqnarray}
where $K_\nu$ is the Macdonald function (modified Bessel function). 

These solutions  $<\rho|\vec
k_{\perp},j,\pm>\,=\,\psi_{\vec k_{\perp},j}^{(\pm)}(\rho)$, for which we
shall use the shorthand notation  
$ \psi_{j}^{(\pm)}(\rho)\equiv\psi_{\vec k_{\perp},j}^{(\pm)}(\rho)$, form a complete set of functions with the
orthonormalization condition \cite{canter}
\begin{eqnarray}
\int^{\infty}_{0}\frac{d\rho}{\rho}\psi^{(\mp)}_{j}(\rho)\psi^{(\mp)}_{j'}(\rho)
= {1\over\pi^2}\int^{\infty}_{0}\frac{d\rho}{\rho}\sqrt{(\mp2{\ii}j-1)\cosh\pi
    j(\mp2{\ii}j'-1)\cosh\pi j'}
K_{{\ii} j\pm\frac{1}{2}}(m\rho)K_{{\ii}j'\pm\frac{1}{2}}(m\rho) = \delta(j-j'). 
\end{eqnarray}

Taking the above formulas and the product of operators (\ref{sq}) into
account, we obtain
\begin{equation}
B_2^2\,=\,4|\Delta|^2+{1\over \rho^2}\left[{1\over
  a}\partial_\eta + {1\over 2}\gamma_{\hat 0}\gamma_{\hat 1}\right
  ]^2-\left [{\partial^2\over \partial \rho^2}+{1\over
  \rho}{\partial\over \partial \rho}-({\vec\gamma} \vec
    \nabla)_{\perp}^2-\sigma^2\right ]
+\left({\mu\over
  a}\right)^2{1\over \rho^2}-2{\mu \over a\rho}[\gamma_{\hat 0}\sigma - {\ii}\gamma_{\hat 0}(\vec\gamma \vec
    \nabla)].
\label{bsqr}
\end{equation} 

For our further calculations,  we also have to transform, in the same
way, the product of operators in $B_1^2$:
\begin{equation}
B_1^2\,=\,
{1\over \rho^2}\left[{1\over a}\partial_\eta+{1\over 2}\gamma_{\hat 0}\gamma_{\hat 1}-{\ii}{\mu\over
    a}\right]^2
-\left [{\partial^2\over \partial \rho^2}+{1\over
  \rho}{\partial\over \partial \rho}-(\vec \gamma \vec
    \nabla)_{\perp}^2-\sigma^2\right ].
\label{b1}
\end{equation}

In order to find nonvanishing condensates $\langle\sigma\rangle$ and
$\langle\Delta^3\rangle$, we should calculate the effective
potential, whose global minimum point provides us with these quantities.
By definition of the effective potential we have
\begin{equation}
V_{\rm eff} = -\frac{S_{\rm eff}}{\int d^{D}x \sqrt{-g}}.
\label{eff}
\end{equation}
In this case, according to (\ref{eff}) we obtain
\begin{eqnarray}
\label{eq.8}
V_{\rm eff}=\frac{3\sigma^2}{2G_1}+\frac
{3\Delta^b\Delta^{\ast b}}{G_2}-\frac{S_q}{\int d^4x\sqrt {-g}}.
\end{eqnarray}
The gap equations correspond to the stationarity condition:
\begin{equation}
 \frac{\partial V_{\rm eff}}{\partial \Delta_0^{3*}}=0,\,\,\frac{\partial V_{\rm eff}}{\partial \sigma}=0.
\label{gap1}
\end{equation}
The gap equations (\ref{gap1}) are obtained by differentiating the
logarithm in (\ref{Sq}) with respect to  $\Delta^{3*}\rm {and}\,\,
\sigma$. Moreover, by taking into account that the position of 
the accelerated observer is defined in (\ref{position}), we can put
$\rho=1/a$.

\section{Chiral symmetry breaking}

First, let us consider a simpler problem, chiral symmetry breaking,
already discussed in \cite{ohsaku2} for the case of a vanishing
chemical potential, and described
qualitatively in the Introduction. In this Section, unlike
in \cite{ohsaku2}, a nonzero chemical potential $\mu$ will be taken
into account. 

Let us first assume that $\Delta=0$ and $\mu=0$. Then according to
(\ref{Sq}), (\ref{sq}),
\[
S_q\,=\,-{3\over 2}{\ii}\tr\log B^2
\]
and the gap equation looks like
\[
\sigma\,=\,-{{\ii}G_1\sigma\over \int d^4x\sqrt {-g}}\tr{1\over
  B^2}.
\]

Now we go over to the momentum representation by replacing
\[
{\ii}{\partial\over \partial \eta}\to k_0, \,\,-{\ii}\vec
    \nabla_{\perp}\to \vec k_{\perp},
\] 
and obtain
\[
\sigma\,=\,-{\ii}G_1\sigma N_f\int{dk_0\over2\pi}\int
{d^2k_{\perp}\over (2\pi)^2}\int_0^{\infty}{d\rho\over\rho}<\vec k_{\perp},k_0,\rho|{1\over
  B^2}|\vec k_{\perp},k_0,\rho>|_{\rho=a^{-1}}, 
\]
where $N_f=2$ is the number of flavors in our problem. 
With the use of the completeness relation (\ref{basis}) of the Rindler basis 
$<\rho|\vec k_{\perp},j,\pm>\,=\,\psi_{\vec k_{\perp},j}^{(\pm)}(\rho)$,
let us go over to this basis in the variable $\rho$, so that $$\rho^2{d^2\over
  d\rho^2}+\rho{d\over d\rho}-m^2\rho^2 \to -(j\pm{{\rm i}\over 2})^2.$$

Next, we are going to an imaginary time coordinate, i.e., to the Euclidean
spacetime in order to consider the thermal effect of 
acceleration~\cite{8a,9}.
The Euclidean Rindler spacetime has a singularity at $\rho=0$,
therefore we have to choose the period of the imaginary time as 
$2\pi/ a$~\cite{brill}.
The Euclidean formalism in Rindler coordinates with a definite period
$\beta=2\pi/ a$ of imaginary time  
coincides with the finite-temperature imaginary time Matsubara
formalism, which is realized by the following substitutions in 
our equations~\cite{finitemp}:
\begin{equation} 
\int\frac{dk_{0}}{2\pi}\to\sum_{n}\frac{1}{\beta}, \quad k_{0}\to {\rm
  i}\omega_{n}, 
\end{equation}
where $\omega_{n}$ is the discrete fermion  frequency defined 
by $\omega_{n}=(2n+1)\pi/\beta$ ($n=0, \pm 1, \pm 2, \cdots$). 

After this, we finally obtain
\begin{equation}
\sigma\,=\,-G_1\sigma N_f\sum_{n}\frac{1}{2\pi}\int\frac{d^2k_{\perp}}{(2\pi)^2}\sum_{\pm}\int_0^{\infty}dj{\left[{1\over\pi}\sqrt{(\mp2{\ii}j-1)\cosh\pi
  j}K_{{\ii}j\pm{1\over 2}}(m a^{-1})\right]^2\over({{\ii}\omega_n\over
  a}-j\pm{\ii})({{\ii}\omega_n\over a}+j)}.
\label{mu0}
\end{equation}

Now, consider the case of a nonzero chemical potential $\mu$. 
Take into consideration that under the operation of charge conjugation 
performed in the above $B^2_{1,2}$ operators one obtains $\mu\to -\mu$. Hence we
should take both signs, $\pm\mu$, into consideration (this is
equivalent to considering also for another branch of the solution of the
Rindler equation with $j\to -j$). Therefore
\[
S_{q1}=-{{\rm i}\over 2}\tr\log B_1^2\to-{{\rm i}\over 2}\tr \log B_{1+}^2 B_{1-}^2
\]
where
\begin{equation}
B_{1\zeta}^2\,=\,
{1\over \rho^2}\left[{1\over a}\partial_\eta +
    {1\over 2}\gamma_{\hat 0}\gamma_{\hat 1}-{\ii}\zeta{\mu\over
    a}\right]^2
-\left [{\partial^2\over \partial \rho^2}+{1\over
  \rho}{\partial\over \partial \rho}-(\vec \gamma \vec
    \nabla)_{\perp}^2-\sigma^2\right ]\quad {\rm with}\quad \zeta=\pm 1.
\label{b11}
\end{equation}
Then, taking the Rindler basis (\ref{solution}), (\ref{basis}) into
 consideration, in the Euclidean spacetime 
 with Matsubara frequencies the above operator takes the form
\begin{equation}
B_{1\zeta}^2\,=\,{1\over \rho^2}\left[{\omega_n\over a}\pm
    {1\over 2}-{\ii}\zeta{\mu\over
    a}\right]^2\,+\,{1\over \rho^2}\left[j\mp{{\ii}\over2}\right]^2.
\label{bsqr2}
\end{equation}
We can write the identity
\[
B_{1+}^2B_{1-}^2\,=\,\left({1\over \rho^2}\left[j\mp{{\ii}\over2}\right]^2+{1\over \rho^2}\left[{\omega_n\over a}\pm
    {1\over 2}-{\ii}{\mu\over a}\right]^2\right)\cdot\left({1\over \rho^2}\left[j\mp{{\ii}\over2}\right]^2+{1\over \rho^2}\left[{\omega_n\over a}\pm
    {1\over 2}+{\ii}{\mu\over a}\right]^2\right)
\]

\begin{equation}
=\left({1\over \rho^2}\left[j\mp{{\ii}\over2}+{\mu\over a}\right]^2+{1\over \rho^2}\left[{\omega_n\over a}\pm
    {1\over 2}\right]^2\right)\cdot\left({1\over \rho^2}\left[j\mp{{\ii}\over2}-{\mu\over a}\right]^2+{1\over \rho^2}\left[{\omega_n\over a}\pm
    {1\over 2}\right]^2\right).
\label{identity}
\end{equation}
From the Bessel equation (\ref{bess}) we have
\[
<\vec k_{\perp},j,\pm|-2\sigma d\sigma\rho^2+2E_jdj|\vec k_{\perp},j,\pm>\,=\,0,
\]
and hence in the case of finite $\mu$ we obtain instead of (\ref{mu0})
\begin{equation}
\sigma\,=\,-G_1\sigma
N_f\sum_{\zeta}\sum_{n}\frac{1}{2\pi}\int\frac{d^2k_{\perp}}{(2\pi)^2}
\sum_{\pm}\int_0^{\infty}dj  
{j\mp{{\ii}\over2}-\zeta{\mu\over a}\over j\mp{{\ii}\over
    2}}\cdot{\left[{1\over\pi}\sqrt{(\mp2{\ii}j-1)\cosh\pi 
  j}K_{{\ii}j\pm{1\over 2}}(m a^{-1})\right]^2\over({{\ii}\omega_n\over
  a}-(j-\zeta{\mu\over a})\pm{\ii})({{\ii}\omega_n\over a}+j-\zeta{\mu\over a})}.
\label{mu}
\end{equation}
Now by summing over Matsubara frequencies $\omega_n=a(n+{1\over 2})$ with the help of the
formula
\begin{equation}
\sum_{n}{1\over ({{\ii}\omega_n\over
  a}-(j-\zeta{\mu\over a})\pm{\ii})({{\ii}\omega_n\over
  a}+j-\zeta{\mu\over a})}=\mp{2\pi{\ii}\over
1\pm2{\ii}(j-\zeta{\mu\over a})}\tanh \pi(j-\zeta{\mu\over a}),
\label{1sum}
\end{equation}
the final result is
obtained
\begin{equation}
\sigma\,=\,-{\ii}G_1\sigma
 N_f \sum_{\zeta}\sum_{\pm}\int\frac{d^2k_{\perp}}{(2\pi)^2}\int_0^{\infty}dj 
\cosh\pi j{\tanh \pi(j-\zeta{\mu\over a})\over \pi^2}(K_{{\ii}j\pm{1\over 2}}(m
  a^{-1}))^2.
\end{equation}
\label{final}

For illustrations, let us estimate the critical acceleration $a_c$, when the
quark condensate vanishes. To this end, we put $\sigma=0$ ($m\to k_{\perp}$) and
integrate over $\vec k_{\perp}$ with the help of the integral
\begin{equation}
\int^{\infty}_{0} k dk K_n(ka^{-1})^2={1\over 2}a^2{n\pi \over \sin n\pi}
\label{integral}
\end{equation}

The final result looks as follows:
\begin{equation}
1={G_1\over 2\pi^2} N_f\sum_{\zeta}\int_0^\infty dq q \tanh
(\pi{q-\zeta\mu\over a}).\label{ffinal}
\end{equation}
The above equation precisely corresponds to the known expression for
the critical curve, obtained for finite temperature and chemical
potential (see, e.g. \cite{klev}), if the correspondence between
the acceleration $a$ and Unruh temperature $T$ is taken into consideration,
$${\pi\over a}\,=\,{1\over 2T}.$$  Now,
recall that the Unruh temperature is given by the relation
\[
T={a\over 2.5\times 10^{22}({\rm cm}\,{\rm s}^{-2})}{\rm K}.
\]
Let us take the value of the maximum critical temperature on the
transition curve for the quark condensate formation $T_m=0.169\,{\rm
  GeV},$ calculated in  \cite{klev}. Then, 
we find for the critical acceleration the following estimate
$a_c=2\pi T=2\pi \times 0.169$ GeV $=3.2\times 10^{35}$ cm/s$^2$. This
value is an order of magnitude larger than the value found for the
case of a vanishing chemical potential  in \cite{ohsaku2}. 

\section{Color symmetry breaking and formation of a diquark condensate}

In order to study the minimum in the variable $\Delta$ of the
diquark condensate, we may here put $\sigma=0$. Now,  we
have
\begin{equation}
S_{q2}=-{\ii}\tr\log B_2^2\,=\,-{\ii}\tr\log[4|\Delta|^2+(-{\rm
  i}\gamma^\nu\nabla_\nu+\mu\gamma^0)({\rm i}\gamma^\mu\nabla_\mu+\mu\gamma^0)].
\label{1B}
\end{equation}
Then after taking into account charge conjugation
\[
S_{q2}\to-{\ii}\tr\log B^2_{2+}B^2_{2-},
\]
where
\[
B^2_{2\zeta}=4|\Delta|^2+(-{\rm i}\gamma^\nu\nabla_\nu+\zeta\mu\gamma^0)({\rm
  i}\gamma^\mu\nabla_\mu+\zeta\mu\gamma^0)
\]
The corresponding gap equation now takes the form
\[
{3\Delta\over G_2}={1\over\int d^4x\sqrt{-g}}{\partial S_{q2}\over\partial \Delta}.
\]
Let us again estimate the value of the critical Unruh temperature
and acceleration, at which the broken color symmetry is restored. For
this purpose, we now put  $\Delta=0$. Then 
the operators in the above equation can be expanded in the Rindler basis
(\ref{basis}), and 
after going over to the Euclidean spacetime and
Matsubara frequencies the gap equation can again be written in the form
\begin{eqnarray}
1 &=& {4\over3} G_2 N_f\sum_{n}\sum_{\zeta}\frac{1}{2\pi}\int\frac{d^{2}k_{\perp}}{(2\pi)^{2}}\int^{\infty}_{0}dj    \nonumber \\
& & \times \Bigl\{ \frac{(-2{\rm i}j-1)\cosh\pi j}{\pi^{2}}(K_{{\rm
    i}j+\frac{1}{2}}(m a^{-1}))^{2}\frac{1}{(({\rm i}\omega_{n}+\zeta\mu)/a-j+{\rm i})
(({\rm i}\omega_{n}-\zeta\mu)/a+j)}  \nonumber \\
& & \qquad
+\frac{(2{\rm i}j-1)\cosh\pi j}{\pi^{2}}(K_{{\rm i}j-\frac{1}{2}}(m
a^{-1}))^{2}\frac{1}{(({\rm i}\omega_{n}+\zeta\mu)/a-j-{\rm i})(({\rm
    i}\omega_{n}-\mu)/a+j)}
\Bigr\} 
\end{eqnarray}
where now 
$m^2=\vec k_{\perp}^2.$

Next, we have to sum over the Matsubara frequencies in the above
equation with the help of the formula (\ref{1sum}). As a result, we obtain
\begin{eqnarray}
1&=& -{4\over3}
\frac{G_2}{2\pi}N_f\int\frac{d^{2}k_{\perp}}{(2\pi)^{2}}\int^{\infty}_{0}dj\sum_{\zeta}2{\ii}\pi
\tanh \pi(j-\zeta{\mu\over a})   \nonumber \\
& & \times \Bigl\{ \frac{(2{\rm i}j+1)\cosh\pi j}{\pi^{2}}(K_{{\rm
    i}j+\frac{1}{2}}(m a^{-1}))^{2}\frac{1}{1+2{\rm i}(j-\zeta\mu/a)}  \nonumber \\
& & \qquad
+\frac{(2{\rm i}j-1)\cosh\pi j}{\pi^{2}}(K_{{\rm i}j-\frac{1}{2}}(m
a^{-1}))^{2}\frac{1}{1-2{\rm i}(j-\zeta\mu/a)}
\Bigr\}.
\end{eqnarray}
Now, the integration over $k$ can be performed with the use of the integral (\ref{integral}).
Then we have
\[
1 = -{\ii} {2\over3(2\pi)^2} G_2a^2N_f\sum_\zeta\int^{\infty}_{0}dj
\tanh\pi(j-\zeta\mu/a)\left[{(1+2{\ii}j)^2\over 1+2{\ii} (j-\zeta\mu/a)}-{(1-2{\ii}j)^2\over 1-2{\ii} (j-\zeta\mu/a)}\right]. 
\]
Changing the integration variable as $j\to q=aj$, and taking into
consideration that in the physical region $q/a\gg~ 1,\,\,(q-~\mu)/a \gg~ 1$, the above
integral can be approximated by
\[
1={2\over 3} {G_2\over \pi^2}N_f\left[\int^{\Lambda}_{0}dq
q^2{\tanh{\pi(q+\mu)\over a}\over q+\mu}\,+\,\int^{\Lambda}_{\mu}dq
q^2{\tanh{\pi(q-\mu)\over a}\over q-\mu}\,+\,\int^{\mu}_{0}dq
q^2{\tanh{\pi(\mu-q)\over a}\over \mu-q}\right],
\]
where the upper limit in the integral was replaced by the 
cutoff $\Lambda$ for the physical regularization. If the
correspondence ${\pi\over a}\,=\,{1\over 2T}$ between
the acceleration $a$ and temperature $T$ is taken into consideration,
this result exactly  
corresponds to the well known formula for the critical curve in the
usual CSC theory at finite temperature \cite{rapp,klev}. Let us again give a
rough estimate of the order of 
the critical acceleration using the numerical results of \cite{klev}.
By taking their value of the critical temperature on the
transition curve for color superconductivity $T_c=40\,{\rm MeV}$, and
the chemical potential $\mu=0.4$ GeV,
we find for the critical acceleration the following estimate
$a_c=2\pi T_c=2\pi \times 0.04$ GeV $=7.5\times 10^{34}$ cm/s$^2$, which
differs from the critical acceleration for restoration of chiral
symmetry by a factor 4.
\section{Summary and Conclusions}
We have investigated the role of the thermalization effect by the
acceleration of an observer for the restoration of chiral
and color symmetries  in quark matter at finite density in
the framework of the NJL model. For this aim, the effective potential and the gap
equations both for the chiral and diquark condensates have been analytically
derived, and on this basis the values of the critical 
acceleration, where the chiral or color
symmetry is restored, have been determined. Obviously, the
acceleration plays here the role of the
temperature, as if the system is placed into a thermostat. In
particular, we have demonstrated that the results
obtained with finite acceleration at the critical points are quite
similar to those with the ordinary thermal situation, describing  phase
transitions with restoration of chiral or color symmetry at finite
temperature and chemical potential.

The dependence 
of chiral and color properties of the quark matter on
the acceleration of the observer may  be useful in the physics of
black holes, where the Rindler 
metric can be considered as an approximation for the description of
the surface gravitational fields\footnote{Of course, massive black
  holes with masses $M\gtrsim M_{\odot}$ have a Hawking temperature
  $T_H\lesssim 10^{-6}$K, which is much smaller than the temperature
  of restoration of symmetry, $T_c\sim 10^{12}$K. On the other
  hand, primordial black holes with masses $ M\sim 10^{12}$kg would lead
  to a temperature $T_H\sim 10^{12}$K of comparable order of
  magnitude.}. Moreover, the investigation of the influence of  strong  
gravitational fields, such as in  compact stars, on the diquark
condensation and thus on the possible existence of color
superconductivity in the core of the compact stars, is also
of great importance. Further investigations in this direction are
under way, and their results will appear in subsequent publications.
\acknowledgments
One of the authors 
(V.Ch.Zh.) gratefully acknowledges the hospitality of
Prof. M.~Mueller-Preussker and his colleagues at the particle theory
group of the Humboldt University extended to him during his stay
there.
This work was supported by DAAD.

\end{document}